\pdfoutput=1
\documentclass[prd,onecolumn,notitlepage,eqsecnum,nofootinbib,floatfix]{revtex4-1}
\usepackage{amssymb}
\usepackage{amsmath}
\usepackage{amsfonts} 
\usepackage{bm}
\allowdisplaybreaks
\begin{document}
\title{Self-force as a probe of global structure} 
\author{Karl Davidson and Eric Poisson}  
\affiliation{Department of Physics, University of Guelph, Guelph,
  Ontario, N1G 2W1, Canada} 
\date{May 2, 2018}
\begin{abstract} 
We calculate the self-force on an electric charge and electric dipole
held at rest in a closed universe that results from joining two copies
of Minkowski spacetime at a common boundary. Spacetime is strictly
flat on each side of the boundary, but there is curvature at the
surface layer required to join the two Minkowski spacetimes. We find
that the self-force on the charge is always directed away from the
surface layer. This is analogous to the case of an electric charge
held at rest inside a spherical shell of matter, for which the
self-force is also directed away from the shell. For the dipole, the
direction of the self-force is a function of the dipole's position and
orientation. Both self-forces become infinite when the charge or
dipole is made to approach the surface layer. This study reveals that
a self-force can arise even when the Riemann tensor vanishes at the
position of the charge or dipole; in such cases the self-force is a
manifestation of the global curvature of spacetime.     
\end{abstract}  
\maketitle

\section{Introduction and summary} 
\label{sec:intro} 

An electric charge held at rest in a curved spacetime creates an
electric field that interacts with the spacetime curvature. In a flat
spacetime the field lines would be isotropically distributed around
the charge, and the field would exert no net force. In the curved
spacetime, however, the isotropy is disturbed, and there is a
self-force acting on the charge \cite{dewitt-brehme:60}.  

The prototypical example of a self-force in curved spacetime
implicates an electric charge $q$ held at a position $r_0$ outside a
nonrotating black hole of mass $M$. This force was first calculated by
Smith and Will \cite{smith-will:80}, building up on earlier work by
a number of researchers \cite{copson:28, cohen-wald:71,
  hanni-ruffini:73, bicak-dvorak:76, linet:76}. It is given by  
\begin{equation} 
\bm{F} = \frac{q^2 M}{r_0^3}\, \bm{n}
\qquad \mbox{(charge outside a black hole)},  
\label{SW}
\end{equation} 
where $\bm{n}$ is a unit vector that points from the black hole to the
charge.\footnote{Throughout the paper we use a Cartesian language for  
vectors. In this case the spacetime is curved and the language is not
entirely appropriate. The equation, however, is still valid. The
vector $\bm{F}$ refers to an orthonormal basis attached to a static
observer at position $r_0$, and $\bm{n}$ is the unit vector that
points radially outward.} The self-force points away from the black
hole, meaning that the external force required to keep the particle in
place is smaller when the particle is charged, compared to what it
would be in the case of a neutral particle. This iconic result was
later generalized to electric charges in the Reissner-Nordstr\"om spacetime
\cite{zelnikov-frolov:82, bini-geralico-ruffini:07}, to scalar charges
\cite{wiseman:00, burko:00b, burko-liu:01}, and to higher-dimensional
black holes \cite{frolov-zelnikov:12a, frolov-zelnikov:12b,
  beach-poisson-nickel:14, taylor-flanagan:15,
  harte-flanagan-taylor:16}. 

Drivas and Gralla \cite{drivas-gralla:11} have shown that the
Smith-Will force of Eq.~(\ref{SW}) is nearly universal, in the sense 
that its expression is nearly independent of the internal composition
of the gravitating body. In other words, two bodies of the same mass
but of distinct internal structure give rise to nearly identical
self-forces, both given approximately by $q^2 M/r_0^3$. There is,
however, a slight dependence on internal structure that produces a 
correction of order $q^2 M^3/r_0^5$, a factor $(M/r_0)^2$ smaller
than the leading-order term; the correction depends on the internal
composition of the gravitating body. The precise nature of these
corrections was identified by Isoyama and Poisson
\cite{isoyama-poisson:12}, who concluded that in principle, the
self-force can be exploited as a probe of internal structure. 

The approximate universality of the Smith-Will force suggests that a
reliable estimate for the self-force acting on a static charge in any
curved spacetime might be that it is given by $q^2$ times a measure of 
the local curvature, as provided by a typical component of the Riemann
tensor in an orthonormal frame. In the case of a spacetime outside a
body of mass $M$, the local curvature at position $r_0$ is measured by 
$M/r_0^3$, and the self-force is indeed given by $q^2 M/r_0^3$.  

The rule of thumb might be adequate in many circumstances, but its
limitations become apparent when the charge is placed in a region of
spacetime in which the local curvature vanishes. An illuminating
example, first examined by Unruh \cite{unruh:76a} and later revisited
by Burko, Liu, and Soen \cite{burko-etal:00}, implicates an
electric charge held at rest at a position $r_0$ inside a spherical
shell of mass $M$ and radius $a$. In this situation, the spacetime
inside the shell is flat, and the local curvature at $r_0$
vanishes. But the spacetime is curved outside the shell, and the
interaction between the electric field and this curvature still gives
rise to a self-force. In this case the self-force provides a probe of
the global curvature of the spacetime. 

The precise expression for the force is complicated --- it is
presented as an infinite sum involving Legendre functions --- but to
leading order in an expansion in powers of $M/a$, it reduces to 
\cite{burko-etal:00} 
\begin{equation} 
\bm{F} \simeq -\frac{q^2 M}{2 a r_0^2} \biggl( 
\frac{r_0/a}{1-r_0^2/a^2} 
+ \frac{1}{2} \ln \frac{1-r_0/a}{1+r_0/a} \biggr)\, \bm{n} 
\qquad (\mbox{charge inside a shell, leading order in $M/a$}), 
\label{sf_shell}
\end{equation} 
where $\bm{n}$ is a unit vector that points toward the charge from the 
center of the shell. Here the self-force points toward the center,
away from the shell, and its scaling with $M$, $a$, and $r_0$ is not
given by a simple expression. When $r_0/a$ is small, the expression
within brackets asymptotes to $\frac{2}{3} (r_0/a)^3$, and 
$\bm{F} \sim  -\frac{1}{3} (q^2 M r_0/a^4)\, \bm{n}$. When $r_0/a$
approaches unity, the expression within brackets is approximately
equal to $[2(1-r_0/a)]^{-1}$, and the self-force becomes
infinite in the limit. This example reveals that the scaling of the
self-force is difficult to estimate when the charge is placed in a
region of vanishing curvature. The rule of thumb proposed previously
is clearly inadequate in such situations.  

In this paper we examine an even more radical example of a self-force 
acting on an electric charge held at rest in a region of vanishing
local curvature. We consider a static, spherically-symmetric spacetime
with a metric given by
\begin{equation} 
ds^2 = -dt^2 + dr^2 + R^2(r)\, d\Omega^2, 
\label{metric} 
\end{equation} 
where $d\Omega^2 := d\theta^2 + \sin^2\theta\, d\phi^2$ and  
\begin{equation} 
R(r) := \left\{ 
\begin{array}{ll} 
r & \qquad 0 \leq r < a \\ 
2a-r & \qquad a \leq r \leq 2a 
\end{array} \right. .
\end{equation} 
The spacetime represents a static universe with closed spatial
sections, which extend from a first center at $r=0$, at which
$R(r)=0$, to a second center at $r=2a$, at which $R(r)$ also
vanishes. The spacetime is strictly flat when $r < a$ and $r > a$, and
it can be thought of as two copies of Minkowski spacetime
joined together at $r = a$. The joint is achieved with a surface
layer, and the Israel junction conditions \cite{israel:66} imply that
the layer possesses a surface mass density $\sigma = (2\pi a)^{-1}$
and (negative) surface pressure $p = -(4\pi a)^{-1}$. The surface
layer has an inertial mass $m = 4\pi a^2 \sigma = 2 a$, and it is the
only place in spacetime where one can find curvature. The Riemann 
tensor is a Dirac distribution supported at $r=a$, and the Einstein
field equations equate its Ricci piece to the distributional
energy-momentum tensor provided by the surface layer. The spacetime is 
admittedly unrealistic, but it nevertheless gives rise to a striking
(and simple) example of a self-force. In this case, the electric field
interacts with curvature that is entirely confined to the surface layer.  

In Sec.~\ref{sec:EM-monopole} we place an electric charge $q$ at a
position $r_0$ in the spacetime, keep it there by means of an external
agent that balances the self-force, calculate the electric field,
observe that it is affected by the global structure of the spacetime,
and find that this field exerts a force on the charge. Taking 
$r_0 < a$ without loss of generality (because the spacetime is
reflection-symmetric across $r = a$), we find that the self-force is
given by     
\begin{equation} 
\bm{F} 
= -\frac{q^2 r_0}{2a^3} \frac{1}{1-(r_0/a)^2}\, \bm{n} 
\qquad (\mbox{charge in the closed universe}).  
\label{self-mono} 
\end{equation} 
The minus sign indicates that the force is directed away from the 
surface layer. It scales as $q^2 r_0/a^3$ when $r_0/a \ll 1$, and it
diverges in the limit $r_0 \to a$.  

A technical complication arises because of the closed spatial sections
of our spacetime. Gauss's law demands that the total charge be zero in
a closed universe, and there must therefore be a second charge $-q$ in
the spacetime. The second charge creates an additional force on the
first charge, and this force must be distinguished from the 
self-force of Eq.~(\ref{self-mono}). This can always be done, because
the self-force depends only on the position $r_0$, while the
interaction force between the two charges depends also on the position
of the second charge. These details are presented in
Sec.~\ref{sec:EM-monopole}. 

The scaling of the self-force with $q^2 r_0/a^3$ and its blow-up at
$r_0 = a$ are reminiscent of the behavior of the self-force in the
case of a charge inside a massive shell; see Eq.~(\ref{sf_shell})
and the discussion that follows. In this case we found the same
blow-up, and a rough scaling with $q^2 M r_0/a^4$. It is
tempting to suggest that the $q^2 r_0/a^3$ scaling is recovered when
$M$ is of the same order of magnitude as $a$. This recovery is 
suggestive, but the suggestion suffers from the drawback that the
surface layer of our closed universe has a vanishing gravitational
mass; partial redemption may come from our earlier observation that
its inertial mass is indeed given by $2a$. Another drawback comes from
the fact that in the example of the massive shell, the scaling with $q^2 M
r_0/a^4$ was identified in a self-force that was valid only to leading
order in an expansion in powers of $M/a$. The self-force, however, was
also calculated in the limit $a \to 2M$ in Ref.~\cite{burko-etal:00}. In
this case it is given by 
\begin{equation} 
\bm{F} \simeq -\frac{q^2 r_0}{a^3} 
\frac{2-r_0^2/a^2}{(1 - r_0^2/a^2)^2}\, \bm{n}
\qquad (\mbox{charge inside a shell, $a \to 2M$}), 
\end{equation} 
and its scaling with $q^2 r_0/a^3$ is confirmed. The blow-up at $r=a$, 
however, is now stronger. 

The spacetime introduced in this paper is sufficiently simple that it
allows an easy investigation of self-forces in unusual and interesting
circumstances. We take advantage of this simplicity and explore a new 
avenue by calculating the self-force acting on an electric dipole
$\bm{p}$ held at rest at position $r_0 < a$ in the spacetime. To the
best of our knowledge, there has been no studies of self-force on a
dipole in curved spacetime, beyond the foundational work found in
Ref.~\cite{messaritaki:07}. We hope that this initial study will
motivate further work in this direction.  

The calculation of the self-force on a dipole presents itself with a
conundrum. The fact that the self-force on a charge $q$ scales with
$q^2$, and is therefore independent of the sign of the charge,
suggests that the forces should simply add up when two opposite
charges are brought together to form a dipole. On the other hand, we
expect that a calculation carried out from first principles would
reveal a self-force that scales as $p^2$, where $p :=
|\bm{p}|$. Because $q = p/\epsilon$, where $\epsilon$ is 
the separation between the charges, the first suggestion would 
produce a force that diverges when $\epsilon \to 0$ with $p$ fixed,
while the second route would produce a finite self-force. The actual
calculation of the self-force shows that it is finite, and the
expectation that the individual self-forces add up is simply wrong.   

In Sec.~\ref{sec:EM-dipole} we calculate the self-force acting on a
point dipole $\bm{p}$ at rest at a position $r=r_0 < a$ in the closed
universe. As in the case of the charge, the dipole is kept in place by
an external agent that balances the self-force. We find that it is
given by    
\begin{equation} 
\bm{F} = -\frac{ p^2r_0}{4a^5} \frac{1}{(1-r_0^2/a^2)^3}\, \bm{f}
\qquad (\mbox{dipole in the closed universe}) 
\label{SF_dipole} 
\end{equation} 
with
\begin{equation} 
\bm{f} := \bigl( 1-r_0^2/a^2 \bigr) (\bm{\hat{p}} \cdot \bm{n})
\bm{\hat{p}}  + \Bigl[ 3-r_0^2/a^2 + 2 \bigl( r_0^2/a^2 \bigr) 
(\bm{\hat{p}} \cdot \bm{n})^2 \Bigr] \bm{n},  
\end{equation} 
where $\bm{\hat{p}} := \bm{p}/p$ and $\bm{n}$ is a unit vector
that points from $r=0$ to the dipole. The directional structure of
the self-force is rich. When $\bm{p}$ is aligned with $\bm{n}$, that
is, when the dipole points in the radial direction, the force is
directed along $-\bm{n}$. When $\bm{p}$ is orthogonal to $\bm{n}$,
that is, when the dipole is transverse to the radial direction, the force
is again directed along $-\bm{n}$. For a generic orientation of the
dipole, the force is directed opposite to a linear combination of
$\bm{\hat{p}}$ and $\bm{n}$. As in the case of a point charge, the
self-force on a dipole diverges when $r_0 \to a$.       
   
\section{Electromagnetic self-force on a point charge} 
\label{sec:EM-monopole} 

We wish to calculate the self-force acting on an electric charge $q$
at rest at $r=r_0$ in the closed universe with the metric of
Eq.~(\ref{metric}). Because the spacetime is reflection symmetric
across $r=a$, there is no loss of generality in taking 
$r_0 < a$. Because the spatial sections are closed, Gauss's law 
demands that the total charge be zero, and therefore there must be a 
second charge $-q$ in the spacetime, which we put at $r=r_1$. We shall
consider the cases $r_1 < a$ and $r_1 > a$. The total force acting on
$+q$ is given by the sum of the self-force and the force exerted by
$-q$.    

The spacetime comes with a timelike Killing vector $t^\alpha$, with
nonvanishing component $t^t = 1$. The vector is covariantly constant,
so that $\nabla_\alpha t_\beta = 0$. 

\subsection{Maxwell's equations} 
\label{subsec:Maxwell} 

Maxwell's equations are $\nabla_{[\alpha} F_{\beta\gamma]} = 0$ and
$\nabla_\beta F^{\alpha\beta} = 4\pi j^\alpha$, where $F_{\alpha\beta}$
is the electromagnetic field tensor, and 
\begin{equation} 
j^\alpha(x) = q \int u_0^\alpha \delta(x,z_0)\, d\tau_0 
- q \int u_1^\alpha \delta(x,z_1)\, d\tau_1 
\end{equation} 
is the current density, with $z^\alpha_0(\tau_0)$ describing the world 
line of the charge $+q$, while $z^\alpha_1(\tau_1)$ describes the world 
line of the charge $-q$; $\tau_0$ and $\tau_1$ are the respective
proper times, $u^\alpha_0 := dz_0^\alpha/d\tau_0$ and 
$u^\alpha_1 := dz_1^\alpha/d\tau_1$ are the respective velocity
vectors, and $\delta(x,z)$ is a scalarized Dirac distribution. The
force exerted on $+q$ is formally given by 
$F^\alpha = q F^\alpha_{\ \beta} u^\beta_0$. This expression must be
regularized to account for the singularity of the field tensor on the
charge's world line.   

For two charges at rest in the spacetime, we have that $u^\alpha_0 
= u^\alpha_1 = t^\alpha$, and the current density is 
$j^\alpha = \rho\, t^\alpha$ with 
\begin{equation} 
\rho = \frac{q}{R_0^2} \delta(r-r_0) \delta(\cos\theta-\cos\theta_0)
\delta(\phi-\phi_0) 
- \frac{q}{R_1^2} \delta(r-r_1) \delta(\cos\theta-\cos\theta_1)
\delta(\phi-\phi_1), 
\label{rho_twocharges} 
\end{equation} 
where $R_0 := R(r_0) = r_0$, $R_1 := R(r_1)$, $(\theta_0,\phi_0)$ are
the polar angles of the charge $+q$, and $(\theta_1,\phi_1)$ are those  
of the charge $-q$. 

We introduce a vector potential $A_\alpha$ and express the field
tensor as $F_{\alpha\beta} = \nabla_\alpha A_\beta  - \nabla_\beta
A_\alpha$. With $A_\alpha = \Phi\, t_\alpha$ and $\Phi$
time-independent, Maxwell's equations become   
\begin{equation} 
\Box \Phi = -4\pi \rho, 
\label{Maxwell0} 
\end{equation} 
where $\Box := g^{\alpha\beta} \nabla_\alpha \nabla_\beta$. The force
acting on $+q$ can then be expressed as  
\begin{equation} 
F_\alpha = q E_\alpha = -q \nabla_\alpha \Phi, 
\label{EMSF} 
\end{equation} 
where $E_\alpha = -\nabla_\alpha \Phi$ is the electric field, related
to the field tensor by $F_{\alpha\beta} = t_\alpha E_\beta - E_\alpha
t_\beta$. 

Incorporating the metric of Eq.~(\ref{metric}), Eq.~(\ref{Maxwell0}) 
reduces to  
\begin{equation} 
\partial_{rr} \Phi + \frac{2R'}{R} \partial_r \Phi 
+ \frac{1}{R^2} D^2 \Phi = -4\pi \rho, 
\label{Maxwell1} 
\end{equation} 
where $R' := dR/dr$ and $D^2 := \partial_{\theta\theta} +
(\cos\theta/\sin\theta) \partial_\theta +
(\sin\theta)^{-2} \partial_{\phi\phi}$ is the Laplacian operator on
the unit two-sphere. The potential $\Phi$ is expanded in spherical
harmonics,  
\begin{equation} 
\Phi(r,\theta,\phi) = \sum_{\ell m} \Phi_{\ell m}(r) 
Y_{\ell m}(\theta,\phi), 
\label{Phi_SH} 
\end{equation} 
and we make use of the completeness relation 
\begin{equation} 
\delta(\cos\theta-\cos\theta') \delta(\phi-\phi') 
= \sum_{\ell m} Y^*_{\ell m}(\theta',\phi') Y_{\ell m}(\theta,\phi),  
\label{SH_completeness} 
\end{equation} 
as well as the eigenvalue equation $D^2 Y_{\ell m} = -\ell(\ell+1)
Y_{\ell m}$. Equation (\ref{Maxwell1}) becomes 
\begin{equation} 
\Phi''_{\ell m} + \frac{2R'}{R} \Phi'_{\ell m} 
- \frac{\ell(\ell+1)}{R^2} \Phi_{\ell m} 
= -\frac{4\pi q}{r_0^2} Y^*_{\ell m}(\theta_0,\phi_0) \delta(r-r_0) 
+ \frac{4\pi q}{R_1^2} Y^*_{\ell m}(\theta_1,\phi_1) \delta(r-r_1). 
\end{equation} 
For simplicity we place the two charges on the same radial line --- so
that $(\theta_1,\phi_1) = (\theta_0,\phi_0)$ --- and for convenience
we align the polar axis with this line, so that $\theta_0 = \theta_1 =
0$. Noting that  
\begin{equation} 
Y^*_{\ell m}(0,\phi') = \sqrt{\frac{2\ell+1}{4\pi}}\, \delta_{m0}, 
\qquad 
Y_{\ell 0}(\theta,\phi) = \sqrt{\frac{2\ell+1}{4\pi}}\, 
P_\ell(\cos\theta),  
\end{equation} 
we find that $\Phi$ is axisymmetric and admits an expansion in
Legendre polynomials,   
\begin{equation} 
\Phi(r,\theta) = \sum_\ell \Phi_\ell(r)\, P_\ell(\cos\theta), 
\label{Phi_Legendre} 
\end{equation} 
with radial functions that satisfy 
\begin{equation} 
\Phi''_\ell + \frac{2R'}{R} \Phi'_\ell 
- \frac{\ell(\ell+1)}{R^2} \Phi_\ell 
= -(2\ell+1) \frac{q}{r_0^2} \delta(r-r_0) 
+ (2\ell+1) \frac{q}{R_1^2} \delta(r-r_1). 
\label{Maxwell2}  
\end{equation} 
These are related by $\Phi_\ell = [(2\ell+1)/(4\pi)]^{1/2}
\Phi_{\ell 0}$ to the radial functions that appear in the original
expansion of Eq.~(\ref{Phi_SH}).  

When $r \neq r_0$ and $r \neq r_1$, Eq.~(\ref{Maxwell2}) admits the
linearly independent solutions $\Phi_\ell = \{ R^\ell, R^{-(\ell+1)} \}$. 
The solution must be regular at $r=0$ and $r=2a$, and to account for
the delta function at $r=r_0$, it must satisfy the junction conditions
\begin{equation} 
\bigl[ \Phi_\ell \bigr]_{r_0} = 0, \qquad 
\bigl[ \Phi'_\ell \bigr]_{r_0} = -(2\ell+1) \frac{q}{r_0^2}, 
\label{junction1} 
\end{equation} 
with $[f]_{r_0} := f(r=r_0^+) - f(r=r_0^-)$ denoting the jump of $f$
across $r=r_0$. Similarly, the junction conditions 
\begin{equation} 
\bigl[ \Phi_\ell \bigr]_{r_1} = 0, \qquad 
\bigl[ \Phi'_\ell \bigr]_{r_1} = (2\ell+1) \frac{q}{R_1^2}  
\label{junction2} 
\end{equation} 
account for the delta function at $r=r_1$. We assume that the surface
layer is electrically inert, so that both $\Phi_\ell$ and $\Phi'_\ell$ are
continuous at $r=a$. 

\subsection{Case $r_1 = 0$ or $r_1 = 2a$} 
\label{subsec:r1=0} 

The simplest situation has the charge $-q$ at either $r_1 = 0$ or
$r_1 = 2a$, the two centers of the spacetime. In this situation
the charge $-q$ creates a monopole field, and its attraction on
$+q$ is simple to describe. With this attraction accounted for, what
is left over is the self-force acting on the original charge.  

When $\ell \neq 0$ the solution to Eq.~(\ref{Maxwell2}) is 
\begin{subequations} 
\begin{align} 
\Phi_\ell(0 < r < r_0) &= q \frac{r^\ell}{r_0^{\ell+1}} 
+ \frac{q}{2\ell} \frac{(r_0 r)^\ell}{a^{2\ell+1}}, \\ 
\Phi_\ell(r_0 < r < a) &= q \frac{r_0^\ell}{r^{\ell+1}} 
+ \frac{q}{2\ell} \frac{(r_0 r)^\ell}{a^{2\ell+1}}, \\ 
\Phi_\ell(a < r < 2a) &= q \frac{(r_0 R)^\ell}{a^{2\ell+1}} 
+ \frac{q}{2\ell} \frac{(r_0 R)^\ell}{a^{2\ell+1}}, 
\end{align} 
\end{subequations} 
where $R = 2a-r$. When $\ell = 0$ the solution is defined up to an
overall additive constant. When $r_1 = 0$ we choose 
\begin{subequations} 
\begin{align} 
\Phi_0(0 < r < r_0) &= -q/r + q/r_0, \\ 
\Phi_0(r_0 < r < a) &= 0, \\ 
\Phi_0(a < r < 2a) &= 0, 
\end{align} 
\end{subequations} 
and note that in this situation, the solution cannot be regular at
$r=0$. When $r_1 = 2a$ we choose instead 
\begin{subequations} 
\begin{align} 
\Phi_0(0 < r < r_0) &= q/r_0, \\ 
\Phi_0(r_0 < r < a) &= q/r, \\ 
\Phi_0(a < r < 2a) &= 2q/a - q/R,  
\end{align} 
\end{subequations} 
and note that this solution cannot be regular at $r=2a$. 

The complete potential in the region $0 \leq r \leq a$ is given by 
\begin{equation} 
\Phi = \Phi^{\rm S} + \Phi^{\rm R} + \Phi^{\rm int}, 
\end{equation} 
where 
\begin{equation} 
\Phi^{\rm S} = q \sum_{\ell=0}^\infty \frac{r_<^\ell}{r_>^{\ell+1}}\, 
  P_\ell(\cos\theta) 
\label{PhiS_modesum}
\end{equation} 
with $r_< := \mbox{min}(r, r_0)$ and $r_> := \mbox{max}(r, r_0)$,
\begin{equation}  
\Phi^{\rm R} = \frac{q}{2} \sum_{\ell=1}^\infty \frac{1}{\ell} 
\frac{(r_0 r)^\ell}{a^{2\ell+1}}\, P_\ell(\cos\theta), 
\label{PhiR_modesum} 
\end{equation} 
and  
\begin{equation} 
\Phi^{\rm int} = \left\{ 
\begin{array}{ll} 
-q/r & \qquad \mbox{$-q$ at $r_1 = 0$} \\  
0 & \qquad \mbox{$-q$ at $r_1 = 2a$}  
\end{array} \right. .
\label{Phiint_mono}
\end{equation} 
The monopole potential $\Phi^{\rm int}$ describes the interaction
between the two charges. When the charge $-q$ is at $r_1=0$, so that
$r_1 < r_0$, the interaction potential is given by $-q/r$, and the
charge $+q$ feels the force created by $-q$, given by $-q^2/r_0^2$;
this is the expected attraction described by the usual Coulomb
law. When, on the other hand, the charge $-q$ is at $r_1 = 2a$, so
that $r_1 > r_0$, the force vanishes because $-q$ represents a
spherical distribution of charge external to the sphere $r=r_0$. 
The potential $\Phi^{\rm S}$ is recognized as the potential that would
be created by the charge $+q$ if it were situated in a globally flat
spacetime. This potential is singular at $r=r_0$, but it produces an
isotropic electric field around the charge, and this field does not
contribute to the self-force acting on the charge. This potential can
be identified with the Detweiler-Whiting {\it singular potential}
\cite{detweiler-whiting:03}. The remaining contribution 
to $\Phi$ is $\Phi^{\rm R}$, which is smooth at $r=r_0$ and is
entirely responsible for the self-force; this is identified with the
Detweiler-Whiting {\it regular potential}.   

The mode sums for $\Phi^{\rm S}$ and $\Phi^{\rm R}$ can be evaluated;
the details are provided in the Appendix. For the singular potential we have
the familiar expression  
\begin{equation} 
\Phi^{\rm S} = \frac{q}{\sqrt{r^2 - 2r_0 r\cos\theta + r_0^2}},  
\label{PhiS_explicit} 
\end{equation} 
and for the regular potential we have  
\begin{equation} 
\Phi^{\rm R} = -\frac{q}{2a} \ln \Biggl( 
\frac{a^2 - r_0 r \cos\theta + \sqrt{r_0^2 r^2 - 2a^2 r_0 r \cos\theta
    + a^4}}{2a^2} \Biggr). 
\label{PhiR_explicit} 
\end{equation} 
The regular potential produces the electric field $E^{\rm R}_a :=
-\partial_a \Phi^{\rm R}$, and the radial component evaluated at
$\theta=0$ is given by 
\begin{equation} 
E_r^{\rm R}(r,\theta=0) = -\frac{qr_0}{2a^3} \frac{1}{1-r_0 r/a^2}. 
\end{equation} 
This expression is valid for $0 \leq r \leq a$ and $r_0 < a$, and the
apparent singularity at $r = a^2/r_0$ is situated beyond $r = a$,
where another (nonsingular) form of solution takes over. The
self-force acting on the charge $+q$ is
$F^{\rm self}_r = q E_r^{\rm R}(r=r_0,\theta=0)$, or 
\begin{equation} 
F^{\rm self}_r = -\frac{q^2 r_0}{2a^3} \frac{1}{1-(r_0/a)^2},  
\label{SF} 
\end{equation} 
as was first displayed in Eq.~(\ref{self-mono}). The expression
applies when $r_0 < a$, and it becomes singular in the limit $r_0 \to
a$. 

\subsection{Case $r_1 < a$} 
\label{subsec:r1<a} 

Next we place the charge $-q$ at an arbitrary position $r_1$ inside
the surface layer, so that $r_1 < a$. For concreteness we present the
calculation assuming that $r_1 > r_0$; the case $r_1 < r_0$ is very
similar, and there is no need to describe it in detail. 

When $\ell \neq 0$, the solution to Eq.~(\ref{Maxwell2}) is given by 
\begin{subequations} 
\begin{align} 
\Phi_\ell(0 < r < r_0) &= q\frac{r^\ell}{r_0^{\ell+1}} 
- q \frac{r^\ell}{r_1^{\ell+1}} 
+ \frac{q}{2\ell} \frac{(r_0 r)^\ell}{a^{2\ell+1}} 
- \frac{q}{2\ell} \frac{(r_1 r)^\ell}{a^{2\ell+1}}, \\ 
\Phi_\ell(r_0 < r < r_1) &= q\frac{r_0^\ell}{r^{\ell+1}} 
- q \frac{r^\ell}{r_1^{\ell+1}} 
+ \frac{q}{2\ell} \frac{(r_0 r)^\ell}{a^{2\ell+1}} 
- \frac{q}{2\ell} \frac{(r_1 r)^\ell}{a^{2\ell+1}}, \\
\Phi_\ell(r_1 < r < a) &= q\frac{r_0^\ell}{r^{\ell+1}} 
- q \frac{r_1^\ell}{r^{\ell+1}} 
+ \frac{q}{2\ell} \frac{(r_0 r)^\ell}{a^{2\ell+1}} 
- \frac{q}{2\ell} \frac{(r_1 r)^\ell}{a^{2\ell+1}}, \\ 
\Phi_\ell(a < r < 2a) &= q\frac{(r_0 R)^\ell}{a^{2\ell+1}} 
- q\frac{(r_1 R)^\ell}{a^{2\ell+1}} 
+ \frac{q}{2\ell} \frac{(r_0 R)^\ell}{a^{2\ell+1}} 
- \frac{q}{2\ell} \frac{(r_1 R)^\ell}{a^{2\ell+1}},  
\end{align} 
\end{subequations} 
where $R = 2a-r$. For $\ell = 0$ we have 
\begin{subequations} 
\begin{align} 
\Phi_0(0 < r < r_0) &= q/r_0 - q/r_1, \\ 
\Phi_0(r_0 < r < r_1) &= q/r - q/r_1, \\
\Phi_0(r_1 < r < a) &= 0, \\ 
\Phi_0(a < r < 2a) &=0. 
\end{align} 
\end{subequations} 
The complete potential in the region $0 \leq r < r_1$ is given by 
\begin{equation} 
\Phi = \Phi^{\rm S} + \Phi^{\rm R} + \Phi^{\rm int}, 
\end{equation} 
with $\Phi^{\rm S}$ given by Eqs.~(\ref{PhiS_modesum}) and
(\ref{PhiS_explicit}), $\Phi^{\rm R}$ given by Eqs.~(\ref{PhiR_modesum}) and
(\ref{PhiR_explicit}), and 
\begin{equation} 
\Phi^{\rm int} = 
-q \sum_{\ell=0}^\infty \frac{r^\ell}{r_1^{\ell+1}}\,
P_\ell(\cos\theta) 
- \frac{q}{2} \sum_{\ell=1}^\infty \frac{1}{\ell} 
\frac{(r_1 r)^\ell}{a^{2\ell+1}}\, P_\ell(\cos\theta). 
\label{Phiint_in_modesum} 
\end{equation} 
A calculation carried out for $r_1 < r_0$ would return the same
expression for $\Phi$, except that the first sum for $\Phi^{\rm int}$
would have $r_1^\ell/r^{\ell+1}$ in front of the Legendre
polynomials; the general expression is $r_<^\ell/r_>^{\ell+1}$, with
$r_< := \mbox{min}(r,r_1)$ and $r_> := \mbox{max}(r,r_1)$. The
potentials keep their interpretation: $\Phi^{\rm S}$ is the singular
potential, $\Phi^{\rm R}$ is the regular potential responsible for the
self-force, and $\Phi^{\rm int}$ describes the interaction between
charges. 

The interaction potential can be evaluated explicitly (see the
Appendix for details). We have 
\begin{equation} 
\Phi^{\rm int} = \frac{-q}{\sqrt{r^2 - 2r_1 r\cos\theta + r_1^2}} 
+ \frac{q}{2a} \ln \Biggl( 
\frac{a^2 - r_1 r \cos\theta + \sqrt{r_1^2 r^2 - 2a^2 r_1 r \cos\theta
    + a^4}}{2a^2} \Biggr). 
\label{Phiint_in_explicit}  
\end{equation} 
The potential produces the electric field $E^{\rm int}_a = -\partial_a
\Phi^{\rm int}$, and the force acting on the charge $+q$ is 
$F^{\rm int}_r = q E^{\rm int}_r(r=r_0,\theta=0)$, or 
\begin{equation} 
F^{\rm int}_r = \pm \frac{q^2}{(r_1-r_0)^2} 
+ \frac{q^2 r_1}{2a^3} \frac{1}{1-r_0 r_1/a^2}, 
\label{intforce_in} 
\end{equation} 
with the positive sign applying when $r_1 > r_0$, and the negative
sign applying when $r_1 < r_0$. Equation (\ref{intforce_in}) reduces to 
$F^{\rm int}_r = -q^2/r_0^2$ when $r_1 = 0$, in agreement with our
results in Sec.~\ref{subsec:r1=0}. The first term in the interaction
force is the usual expression of Coulomb's law; the force is directed
toward the $-q$ charge at $r=r_1$. The second term is a modification
to Coulomb's law contributed by the global curvature of the
spacetime; it is directed toward the surface layer. 

\subsection{Case $r_1 > a$} 
\label{subsec:r1>a} 

In this section we place the charge $-q$ at an arbitrary position
$r_1$ outside the surface layer, so that $r_1 > a$. When 
$\ell \neq 0$, the solution to Eq.~(\ref{Maxwell2}) is given by  
\begin{subequations} 
\begin{align} 
\Phi_\ell(0 < r < r_0) &= q\frac{r^\ell}{r_0^{\ell+1}} 
- q \frac{(R_1 r)^\ell}{a^{2\ell+1}} 
+ \frac{q}{2\ell} \frac{(r_0 r)^\ell}{a^{2\ell+1}} 
- \frac{q}{2\ell} \frac{(R_1 r)^\ell}{a^{2\ell+1}}, \\ 
\Phi_\ell(r_0 < r < a) &= q\frac{r_0^\ell}{r^{\ell+1}} 
- q \frac{(R_1 r)^\ell}{a^{2\ell+1}} 
+ \frac{q}{2\ell} \frac{(r_0 r)^\ell}{a^{2\ell+1}} 
- \frac{q}{2\ell} \frac{(R_1 r)^\ell}{a^{2\ell+1}}, \\
\Phi_\ell(a < r < r_1) &= q\frac{(r_0 R)^\ell}{a^{2\ell+1}} 
- q \frac{R_1^\ell}{R^{\ell+1}} 
+ \frac{q}{2\ell} \frac{(r_0 R)^\ell}{a^{2\ell+1}} 
- \frac{q}{2\ell} \frac{(R_1 R)^\ell}{a^{2\ell+1}}, \\ 
\Phi_\ell(r_1 < r < 2a) &= q\frac{(r_0R)^\ell}{a^{2\ell+1}} 
- q\frac{R^\ell}{R_1^{\ell+1}} 
+ \frac{q}{2\ell} \frac{(r_0 R)^\ell}{a^{2\ell+1}} 
- \frac{q}{2\ell} \frac{(R_1 R)^\ell}{a^{2\ell+1}},  
\end{align} 
\end{subequations} 
where $R = 2a-r$ and $R_1 = 2a-r_1$. For $\ell = 0$ we have 
\begin{subequations} 
\begin{align} 
\Phi_0(0 < r < r_0) &= q/r_0 - q/a, \\ 
\Phi_0(r_0 < r < a) &= q/r - q/a, \\
\Phi_0(a < r < r_1) &= -q/R + q/a, \\ 
\Phi_0(r_1 < r < 2a) &=-q/R_1 + q/a. 
\end{align} 
\end{subequations} 
The complete potential in the region $0 \leq r \leq a$ is given by 
\begin{equation} 
\Phi = \Phi^{\rm S} + \Phi^{\rm R} + \Phi^{\rm int}, 
\end{equation} 
with $\Phi^{\rm S}$ given by Eqs.~(\ref{PhiS_modesum}) and
(\ref{PhiS_explicit}), $\Phi^{\rm R}$ given by Eqs.~(\ref{PhiR_modesum}) and
(\ref{PhiR_explicit}), and with $\Phi^{\rm int}$ now given by 
\begin{equation} 
\Phi^{\rm int} = 
-q \sum_{\ell=0}^\infty \frac{(R_1 r)^\ell}{a^{2\ell+1}}\,
P_\ell(\cos\theta) 
- \frac{q}{2} \sum_{\ell=1}^\infty \frac{1}{\ell} 
\frac{(R_1 r)^\ell}{a^{2\ell+1}}\, P_\ell(\cos\theta). 
\label{Phiint_out_modesum} 
\end{equation} 

The interaction potential is given explicitly by (see Appendix) 
\begin{equation} 
\Phi^{\rm int} = \frac{-qa}{\sqrt{R_1^2 r^2 - 2a^2 R_1 r\cos\theta + a^4}} 
+ \frac{q}{2a} \ln \Biggl( 
\frac{a^2 - R_1 r \cos\theta + \sqrt{R_1^2 r^2 - 2a^2 R_1 r \cos\theta
    + a^4}}{2a^2} \Biggr), 
\label{Phiint_out_explicit}  
\end{equation} 
and it produces the electric field $E^{\rm int}_a = -\partial_a
\Phi^{\rm int}$. The force acting on the charge $+q$ is 
$F^{\rm int}_r = q E^{\rm int}_r(r=r_0,\theta=0)$, or 
\begin{equation} 
F^{\rm int}_r = \frac{q^2 R_1}{2a^3} 
\frac{3 - r_0 R_1/a^2}{(1 - r_0 R_1/a^2)^2}. 
\label{intforce_out} 
\end{equation} 
The force vanishes when $R_1 = 0$, or $r_1 = 2a$, in agreement with
our results in Sec.~\ref{subsec:r1=0}. With the charges situated on
opposite sides of the surface layer, the interaction force bears
little resemblance to the usual Coulomb force, and it is always
directed toward the surface layer.   

\section{Electromagnetic self-force on a point dipole} 
\label{sec:EM-dipole} 

In this section we consider a point electric dipole $\bm{p}$ situated
at $r=r_0 < a$ (inside the surface layer), and we calculate the 
self-force on this dipole. 

\subsection{Point dipole} 

Because the dipole is at rest in a local patch of Minkowski spacetime,
it is convenient to describe its local physics in a Newtonian language
involving Cartesian vectors such as $\bm{p}$; relativistic aspects
reveal themselves only when we integrate Maxwell's equations for the
electrostatic potential $\Phi$. We shall use both the original
spherical coordinates $(r,\theta,\phi)$ and the associated Cartesian
coordinates $(x,y,z)$, but express all vectors and tensors in
Cartesian coordinates. We recall that the vector basis attached to 
the spherical coordinates is given by  
\begin{subequations} 
\begin{align} 
\bm{\hat{r}} &= \sin\theta\cos\phi\, \bm{\hat{x}} 
+ \sin\theta\sin\phi\, \bm{\hat{y}} 
+ \cos\theta\, \bm{\hat{z}}, \\ 
\bm{\hat{\theta}} &= \cos\theta\cos\phi\, \bm{\hat{x}} 
+ \cos\theta\sin\phi\, \bm{\hat{y}} 
- \sin\theta\, \bm{\hat{z}}, \\ 
\bm{\hat{\phi}} &= -\sin\phi\, \bm{\hat{x}} 
+ \cos\phi\, \bm{\hat{y}},   
\end{align} 
\end{subequations} 
in terms of the Cartesian basis vectors. All vectors have a unit
length. 

The charge density of a point dipole is given by 
\begin{equation} 
\rho(\bm{x}) = -\bm{p} \cdot \bm{\nabla} \delta(\bm{x}-\bm{r}_0),  
\label{rho_dipole} 
\end{equation} 
where $\delta(\bm{x}-\bm{r}_0)$ is a three-dimensional delta function,
and $\bm{r}_0 = (0,0,r_0)$ is the dipole's position vector. The
potential $\Phi$ created by the dipole satisfies Eq.~(\ref{Maxwell0}),
or its explicit expression of Eq.~(\ref{Maxwell1}). The force exerted
on the dipole is given formally by 
\begin{equation} 
F_a = p^b \partial_b E_a = -p^b \partial_{ab} \Phi,  
\label{force_dipole} 
\end{equation} 
in which $E_a := -\partial_a \Phi$ is the electric field, and the
potential is evaluated at $\bm{x}=\bm{r}_0$ after
differentiation. This expression must be regularized before a
meaningful result is obtained for the self-force.  

\subsection{Dipole in the $z$ direction}  
\label{subsec:zdipole} 

We first calculate the potential for a dipole aligned with the
$z$ axis. We have $\bm{p} = p\, \bm{\hat{z}}$, so that 
$p \equiv p_z$. We wish to perform the calculation in spherical
coordinates, and to handle the coordinate singularity on the polar
axis --- $\phi$ is not defined there --- we first place the dipole at
$(r_0, \theta_0, \phi_0)$ and eventually take the limit
$\theta_0 \to 0$; the limit is independent of $\phi_0$. We also put
the dipole in the direction of $\bm{\hat{r}}_0$, the radial unit
vector evaluated at $(\theta=\theta_0, \phi=\phi_0)$; in the limit the
dipole becomes aligned with the $z$ axis. 

Working momentarily in Cartesian coordinates, the dipole's charge
density is given by 
\begin{equation} 
\rho = -p \bm{\hat{r}_0} \cdot \bm{\nabla} \delta(\bm{x}-\bm{r}_0) 
= +p \bm{\hat{r}_0} \cdot \bm{\nabla}_0 \delta(\bm{x}-\bm{r}_0), 
\end{equation} 
where $\bm{\nabla}_0$ is the gradient operator associated with the  
variables contained in $\bm{r}_0$. Switching now to the spherical
coordinates, we have  
\begin{equation} 
\rho = p \frac{\partial}{\partial r_0} 
\biggl[ \frac{\delta(r-r_0)}{r^2} \delta(\cos\theta-\cos\theta_0) 
\delta(\phi-\phi_0) \biggr], 
\end{equation}   
or 
\begin{equation} 
\rho = -p \biggl[ \frac{\delta'(r-r_0)}{r_0^2} 
+ 2\frac{\delta(r-r_0)}{r_0^3} \biggr] \delta(\cos\theta-\cos\theta_0)  
\delta(\phi-\phi_0), 
\label{rho-rdipole} 
\end{equation} 
with a prime indicating differentiation with respect to $r$; we made
use of the distributional identity $f(r) \delta'(r-r_0) = f(r_0)
\delta'(r-r_0) - f'(r_0) \delta(r-r_0)$. The same expression for
$\rho$ can be obtained from the two-charge model of
Eq.~(\ref{rho_twocharges}) by letting $r_1 = r_0 - \delta r$,
$\theta_1 = \theta_0$, $\phi_1 = \phi_0$, and taking the limit 
$\delta r \to 0$ with $p := q\delta r$ kept fixed. 

The calculation of $\Phi$ proceeds as in
Sec.~\ref{subsec:Maxwell}. In the limit $\theta_0 \to 0$ with
$\phi_0$ arbitrary, the potential is axisymmetric and admits the
decomposition   
\begin{equation} 
\Phi(r,\theta) = \sum_\ell \Phi_\ell(r)\, P_\ell(\cos\theta), 
\end{equation} 
with radial functions that satisfy 
\begin{equation} 
\Phi''_\ell + \frac{2R'}{R} \Phi'_\ell 
- \frac{\ell(\ell+1)}{R^2} \Phi_\ell 
= (2\ell+1) p \biggl[ \frac{\delta'(r-r_0)}{r_0^2} 
+ 2\frac{\delta(r-r_0)}{r_0^3} \biggr].  
\end{equation} 
When $r \neq r_0$ the differential equation admits the linearly
independent solutions $\Phi_\ell = \{ R^\ell, R^{-(\ell+1)} \}$. 
The solution must be regular at $r=0$ and $r=2a$, and to account for
the singularity at $r=r_0$, it must satisfy the junction conditions
\begin{equation} 
\bigl[ \Phi_\ell \bigr]_{r_0} = (2\ell+1) \frac{p}{r_0^2}, \qquad  
\bigl[ \Phi'_\ell \bigr]_{r_0} = 0. 
\end{equation} 
We again assume that the surface layer is electrically inert, so that
both $\Phi_\ell$ and $\Phi'_\ell$ are continuous at $r=a$. 

The solution that satisfies all these requirements is 
\begin{subequations} 
\begin{align} 
\Phi_\ell(0 < r < r_0) &= -(\ell+1) p \frac{r^\ell}{r_0^{\ell+2}} 
+ \frac{1}{2} p \frac{r_0^{\ell-1} r^\ell}{a^{2\ell+1}}, \\ 
\Phi_\ell(r_0 < r < a) &= \ell p \frac{r_0^{\ell-1}}{r^{\ell+1}} 
+ \frac{1}{2} p \frac{r_0^{\ell-1} r^\ell}{a^{2\ell+1}}, \\  
\Phi_\ell(a < r < 2a) &= \frac{1}{2}(2\ell+1) p 
\frac{r_0^{\ell-1} R^\ell}{a^{2\ell+1}}. 
\end{align} 
\end{subequations} 
The complete potential in the region $0 < r < a$ is then 
\begin{equation} 
\Phi = \Phi^{\rm S} + \Phi^{\rm R} 
\end{equation} 
with 
\begin{equation} 
\Phi^{\rm S} = p \frac{\partial}{\partial r_0} 
\sum_{\ell=0}^\infty \frac{r_<^\ell}{r_>^{\ell+1}} P_\ell(\cos\theta)  
\end{equation} 
and 
\begin{equation} 
\Phi^{\rm R} = \frac{1}{2} p \sum_{\ell=0}^\infty 
\frac{r_0^{\ell-1} r^\ell}{a^{2\ell+1}} P_\ell(\cos\theta), 
\end{equation} 
where $r_< := \mbox{min}(r,r_0)$ and $r_> := \mbox{max}(r,r_0)$. 

Comparison with Eqs.~(\ref{PhiS_modesum}) and (\ref{PhiS_explicit})
allows us to evaluate the sums, and we arrive at 
\begin{equation} 
\Phi^{\rm S} 
= \frac{p(r\cos\theta - r_0)}{(r^2 - 2r_0 r \cos\theta + r_0^2)^{3/2}} 
\label{PhiS_zdipole} 
\end{equation} 
and 
\begin{equation} 
\Phi^{\rm R} = \frac{pa}{2r_0 s^2}, 
\label{PhiR_zdipole} 
\end{equation} 
where 
\begin{equation} 
s^2 := \bigl( r_0^2 r^2 - 2a^2 r_0 r\cos\theta + a^4 \bigr)^{1/2}. 
\label{s2_def} 
\end{equation} 
Equation (\ref{PhiS_zdipole}) is the familiar expression for the
potential of a point dipole aligned with the $z$ axis, when the
dipole is placed in a globally flat spacetime; this potential diverges
at the dipole's position, and it can be identified with the singular
Detweiler-Whiting potential \cite{detweiler-whiting:03}. The potential
of Eq.~(\ref{PhiR_zdipole}) is smooth at the dipole's position, and it
can be identified with the regular Detweiler-Whiting potential. The
self-force acting on the dipole will come entirely from
$\Phi^{\rm R}$.  

\subsection{Dipole in the $x$ direction} 
\label{subsec:xdipole} 

Next we take the dipole to be aligned with the $x$ axis, so that
$\bm{p} = p\, \bm{\hat{x}}$ and $p \equiv p_x$. To set up the
calculation we place the dipole at $(r_0, \theta_0, \phi_0)$
and align it with $\bm{\hat{\theta}}_0$, the angular unit vector
evaluated at the dipole's position. Taking the limit 
$\theta_0 \to 0$ with $\phi_0 = 0$ will take the dipole to the
$z$ axis and align it with the $x$ direction. 

The charge density is given by 
\begin{equation} 
\rho = -p \bm{\hat{\theta}}_0 \cdot \bm{\nabla} 
\delta(\bm{x} - \bm{r}_0)
= p \bm{\hat{\theta}}_0 \cdot \bm{\nabla}_0  
\delta(\bm{x} - \bm{r}_0)
= \frac{p}{r_0} \frac{\partial}{\partial \theta_0}  
\biggl[ \frac{\delta(r-r_0)}{r^2} \delta(\cos\theta-\cos\theta_0)
\delta(\phi-\phi_0) \biggr], 
\end{equation} 
or 
\begin{equation} 
\rho = \frac{p}{r_0^3} \delta(r-r_0) \partial_{\theta_0}  
\delta(\cos\theta-\cos\theta_0) \delta(\phi-\phi_0). 
\end{equation} 
The same expression can be obtained from the two-charge model of
Eq.~(\ref{rho_twocharges}) by letting $r_1 = r_0$, $\theta_1 =
\theta_0 - \delta\theta$, $\phi_1 = \phi_0$, and taking the limit
$\delta\theta \to 0$ keeping $p := q r_0 \delta\theta$ fixed. 

The potential is expanded in spherical harmonics as in
Eq.~(\ref{Phi_SH}), and the completeness relation of
Eq.~(\ref{SH_completeness}) is differentiated with respect to
$\theta_0$. The identity 
\begin{equation} 
\partial_\theta Y_{\ell m} 
= \frac{1}{2} \sqrt{(\ell-m)(\ell+m+1)} Y_{\ell,m+1} e^{-i\phi} 
- \frac{1}{2} \sqrt{(\ell+m)(\ell-m+1)} Y_{\ell,m-1} e^{i\phi} 
\end{equation} 
allows us to express $\partial_{\theta_0} Y_{\ell m}(\theta_0,\phi_0)$
in terms of spherical harmonics, and to take the limit
$\theta_0 \to 0$ with $\phi_0 = 0$; we obtain   
\begin{equation} 
\lim_{\theta_0 \to 0} 
\frac{\partial}{\partial \theta_0} Y_{\ell m}(\theta_0, 0)  
= \frac{1}{2} \sqrt{\frac{2\ell+1}{4\pi}} \sqrt{\ell(\ell+1)}
\bigl( \delta_{m,-1} - \delta_{m,1} \bigr).
\end{equation} 
This relation implies that $\Phi_{\ell,-1} = -\Phi_{\ell,1}$. With 
\begin{equation} 
Y_{\ell,\pm 1} = \mp \sqrt{\frac{2\ell+1}{4\pi}} 
\frac{1}{\sqrt{\ell(\ell+1)}} P_\ell^1(\cos\theta) e^{\pm i\phi},  
\end{equation} 
where $P_\ell^1$ is an associated Legendre function, and the
definition  
\begin{equation} 
\Phi_\ell := -2 \sqrt{\frac{2\ell+1}{4\pi}}
\frac{1}{\sqrt{\ell(\ell+1)}} \Phi_{\ell,1}, 
\end{equation} 
we find that the potential admits the decomposition 
\begin{equation} 
\Phi(r,\theta,\phi) = \sum_{\ell=1}^\infty \Phi_\ell(r) 
P_\ell^1(\cos\theta) \cos\phi 
\end{equation} 
with radial functions that satisfy 
\begin{equation} 
\Phi''_\ell + \frac{2R'}{R} \Phi'_\ell 
- \frac{\ell(\ell+1)}{R^2} \Phi_\ell 
= -(2\ell+1) \frac{p}{r_0^3} \delta(r-r_0). 
\end{equation} 

The solution to the differential equation is 
\begin{subequations}
\begin{align} 
\Phi_\ell(0 < r < r_0) &= p \frac{r^\ell}{r_0^{\ell+2}} 
+ \frac{p}{2\ell} \frac{r_0^{\ell-1} r^\ell}{a^{2\ell+1}}, \\ 
\Phi_\ell(r_0 < r < a) &= p \frac{r_0^{\ell-1}}{r^{\ell+1}} 
+ \frac{p}{2\ell} \frac{r_0^{\ell-1} r^\ell}{a^{2\ell+1}}, \\ 
\Phi_\ell(a < r < 2a) &= p \frac{r_0^{\ell-1} R^\ell}{a^{2\ell+1}} 
+ \frac{p}{2\ell} \frac{r_0^{\ell-1} R^\ell}{a^{2\ell+1}}, 
\end{align}
\end{subequations} 
and the potential in the region $0 < r < a$ is 
\begin{equation} 
\Phi = \Phi^{\rm S} + \Phi^{\rm R} 
\end{equation} 
with 
\begin{equation} 
\Phi^{\rm S} = \frac{p}{r_0} \sum_{\ell=1}^\infty 
\frac{r_<^\ell}{r_>^{\ell+1}} P_\ell^1(\cos\theta) \cos\phi 
\end{equation} 
and 
\begin{equation} 
\Phi^{\rm R} = \frac{1}{2} p \sum_{\ell=1}^\infty 
\frac{1}{\ell} \frac{r_0^{\ell-1} r^\ell}{a^{2\ell+1}} 
P_\ell^1(\cos\theta) \cos\phi. 
\end{equation} 
The sum for $\Phi^{\rm S}$ can be evaluated by noting that
$P_\ell^1(\cos\theta) = -(d/d\theta) P_\ell(\cos\theta)$, and we obtain 
\begin{equation} 
\Phi^{\rm S} = \frac{p r \sin\theta \cos\phi}{(r^2-2r_0 r\cos\theta 
+ r_0^2)^{3/2}}. 
\end{equation} 
This is the familiar expression for the potential of a point dipole in
a globally flat spacetime, when the dipole is aligned with the $x$
axis; this potential diverges at the dipole's position, and it can be
identified with the singular Detweiler-Whiting potential
\cite{detweiler-whiting:03}. The sum for $\Phi^{\rm R}$ can also be
evaluated by comparing with Eqs.~(\ref{PhiR_modesum}) and
(\ref{PhiR_explicit}). After differentiation with respect to $\theta$
we arrive at  
\begin{equation} 
\Phi^{\rm R} = \frac{p}{2a} 
\frac{ (a^2+s^2) r\sin\theta\cos\phi }
{ s^2(a^2-r_0 r \cos\theta +s^2) }, 
\label{PhiR_xdipole} 
\end{equation} 
where $s^2$ is defined by Eq.~(\ref{s2_def}). This potential is smooth
at the dipole's position, and it can be identified with the regular
Detweiler-Whiting potential. The self-force acting the dipole will
originate entirely from $\Phi^{\rm R}$.  

\subsection{Dipole in the $y$ direction} 
\label{subsec:ydipole} 

The calculation of $\Phi$ for a dipole aligned with the $y$ axis
proceeds as in Sec.~\ref{subsec:xdipole}. The steps are identical,
except for the fact that the limit $\theta_0 \to 0$ is now taken
with $\phi_0 = \pi/2$ to produce the correct orientation for the
dipole. The end result for $\Phi^{\rm R}$ is 
\begin{equation} 
\Phi^{\rm R} = \frac{p}{2a} 
\frac{ (a^2+s^2) r\sin\theta\sin\phi }
{ s^2(a^2-r_0 r \cos\theta +s^2) }, 
\label{PhiR_ydipole} 
\end{equation} 
which can be obtained directly from Eq.~(\ref{PhiR_xdipole}) by
replacing the factor of $\cos\phi$ with $\sin\phi$. 

\subsection{Self-force on a dipole}  

The complete $\Phi^{\rm R}$ for an arbitrarily aligned dipole can be 
obtained by combining Eqs.~(\ref{PhiR_zdipole}), 
(\ref{PhiR_xdipole}), and (\ref{PhiR_ydipole}). We get
\begin{equation} 
\Phi^{\rm R} = \frac{(a^2+s^2) ( p_x r\sin\theta\cos\phi 
+ p_y r\sin\theta\sin\phi ) }{2a s^2 (a^2-r_0 r\cos\theta + s^2)}  
+ \frac{a p_z}{2r_0 s^2}, 
\end{equation} 
with $s^2$ defined by Eq.~(\ref{s2_def}). The self-force acting on the
dipole is calculated from Eq.~(\ref{force_dipole}), which we write
as  
\begin{equation} 
F^{\rm self}_a = -p^b \partial_{ab} \Phi^{\rm R}.  
\end{equation} 
To calculate the components of the force we express the potential in
Cartesian coordinates, take the derivatives, and evaluate the result
at $x = y = 0$ and $z = r_0$. We obtain 
\begin{subequations} 
\begin{align} 
F^{\rm self}_x &= -\frac{ r_0 p_x p_z }{4 a^5 (1-r_0^2/a^2)^2}, \\ 
F^{\rm self}_y &= -\frac{ r_0 p_y p_z }{4 a^5 (1-r_0^2/a^2)^2}, \\ 
F^{\rm self}_z &= -\frac{r_0 \bigl[ (3-r_0^2/a^2) (p_x^2 + p_y^2) 
  + 4 p_z^2\bigr]}{ 4a^5(1 - r_0^2/a^2)^3 }. 
\end{align}
\end{subequations} 
This can be put in the vectorial form of Eq.~(\ref{SF_dipole}), with 
$\bm{n} := \bm{r}_0/r_0 = \bm{\hat{z}}$.   

\begin{acknowledgments} 
This work was supported by the Natural Sciences and Engineering
Research Council of Canada.  
\end{acknowledgments} 

\appendix 
\section{Derivation of Eqs.~(\ref{PhiS_explicit}),
  (\ref{PhiR_explicit}), (\ref{Phiint_in_explicit}), and
  (\ref{Phiint_out_explicit})}  

Equation (\ref{PhiS_explicit}) follows directly from
Eq.~(\ref{PhiS_modesum}) and the identity 
\begin{equation} 
(1-2tx+t^2)^{-1/2} = \sum_{\ell=0}^\infty t^\ell P_\ell(x),  
\label{generating} 
\end{equation} 
which holds for $|t| < 1$. 

To obtain Eq.~(\ref{PhiR_explicit}), we note that
Eq.~(\ref{PhiR_modesum}) can be expressed as 
\begin{equation} 
\Phi^{\rm R} = \frac{qa}{2} \int dr_0 \Biggl[ 
\frac{1}{r_0^2} \sum_{\ell=1}^\infty
\frac{r^\ell}{(a^2/r_0)^{\ell+1}}\, P_\ell(\cos\theta) \Biggr],  
\end{equation} 
or 
\begin{equation} 
\Phi^{\rm R} = \frac{qa}{2} \int dr_0 \Biggl[ 
\frac{1}{r_0^2} \sum_{\ell=0}^\infty
\frac{r^\ell}{(a^2/r_0)^{\ell+1}}\, P_\ell(\cos\theta) 
- \frac{1}{a^2 r_0} \Biggr].  
\end{equation} 
Making use of Eq.~(\ref{generating}), the sum over Legendre
polynomials evaluates to 
\begin{equation} 
\sum_{\ell=0}^\infty
\frac{r^\ell}{(a^2/r_0)^{\ell+1}}\, P_\ell(\cos\theta) 
=\frac{r_0}{\sqrt{r_0^2 r^2 - 2 a^2 r_0 r\cos\theta + a^4}}, 
\end{equation} 
and the potential becomes 
\begin{equation} 
\Phi^{\rm R} = \frac{qa}{2} \int dr_0 \Biggl[ 
\frac{1}{r_0 \sqrt{r_0^2 r^2 - 2 a^2 r_0 r\cos\theta + a^4}} 
- \frac{1}{a^2 r_0} \Biggr] 
\end{equation} 
Direct integration gives rise to Eq.~(\ref{PhiR_explicit}) after
adjusting the constant of integration so that 
$\Phi^{\rm R}(r = 0,\theta) = 0$, as implied by
Eq.~(\ref{PhiR_modesum}).  

The steps required to arrive at Eq.~(\ref{Phiint_in_explicit}) from
Eq.~(\ref{Phiint_in_modesum}) are identical to those described
previously, with the changes $q \to -q$ and $r_0 \to r_1$. To go from
Eq.~(\ref{Phiint_out_modesum}) to Eq.~(\ref{Phiint_out_explicit}) we
let $q' := -qa/R_1$ and $R_1' := a^2/R_1$ in the first sum, which
becomes 
\begin{equation} 
q' \sum_{\ell=0}^\infty \frac{r^\ell}{(R'_1)^{\ell+1}}\,
P_\ell(\cos\theta) 
= \frac{q'}{\sqrt{r^2 - 2 R_1' r\cos\theta + R_1^{\prime 2}}}, 
\end{equation} 
and which gives rise to the first term in
Eq.~(\ref{Phiint_out_explicit}). It is interesting to note that the 
expressions for $q'$ and $R'_1$ are the same ones that arise for the
image charge in the problem of a point charge outside a grounded,
spherical conductor. The second sum in Eq.~(\ref{Phiint_out_modesum})
is of the same form as Eq.~(\ref{PhiR_modesum}) with $r_0 \to R_1$,
and it leads directly to the second term in
Eq.~(\ref{Phiint_out_explicit}).   

\bibliography{../bib/master}
\end{document}